\documentclass[fleqn,usenatbib]{mnras}

\usepackage[english]{babel}
\usepackage[utf8x]{inputenc}
\usepackage[T1]{fontenc}
\usepackage{ae,aecompl}
\usepackage{amsmath}
\usepackage{amssymb}
\usepackage{graphicx}
\usepackage[colorinlistoftodos]{todonotes}
\usepackage[para]{threeparttable}

\title{Slingshot prominence evolution for a solar-like star}
\author[Villarreal D'Angelo et al.]{
Carolina Villarreal D'Angelo $^{1,2}$\thanks{E-mail: villarrc@tcd.ie}, Moira Jardine $^{1}$, Colin P. Johnstone $^{3}$  \& Victor See$^{4}$
\\
$^{1}$SUPA, School of Physics and Astronomy, University of St Andrews, North Haugh, KY16 9SS, UK\\
$^{2}$School of Physics, Trinity College Dublin, College Green, Dublin 2, Ireland \\
$^{3}$Department of Astrophysics, University of Vienna, Türkenschanzstrasse 17, 1180 Vienna, Austria\\
$^{4}$University of Exeter, Department of Physics and Astronomy, Stocker Road, Devon, Exeter, EX4 4QL, UK \\
}

\date{Accepted 2019 February 13. Received 2019 February 12; in original form 2018 August 20}

\pubyear{2019}
\begin{document}
\label{firstpage}
\pagerange{\pageref{firstpage}--\pageref{lastpage}}
\maketitle

\begin{abstract}
Although the present-day Sun rotates too slowly to exhibit centrifugally-supported ``slingshot prominences'', at some time during its past it may have formed these clouds of cool gas and ejected them into the interplanetary medium. We determine the time period for this behaviour by using a rotation evolution code to derive the properties of the formation and ejection of slingshot prominences during the lifetime of a star similar to our Sun. The mass, mass loss rate and rate of ejection of these prominences are calculated using the analytical expression derived in our previous work. We find that for stars with an initial rotation rate larger than $4.6\, \Omega_\odot$, about half of all solar mass stars, slingshot prominences will be present even after the star reaches the main sequence phase. In a fast rotator, this means that prominences can form until the star reaches $\sim 800$ Myr old. 
Our results also indicate that the mass and lifetime of this type of prominence have maximum values when the star reaches the ZAMS at an age of $\sim 40$ Myr for a solar mass star. 
\end{abstract}

\section{Introduction}
Since the discovery of slingshot prominences in the fast-rotating cool-dwarf AB Dor \citep{collier1989I,collier1989II}, new examples have been found in both single and binary stars \citep{Barnes2000,Barnes2001,Byrne1996,collier1992,Dunstone2006,Donati2000,Eibe1998,Hall1992,Kolbin2017,Skelly2008,Skelly2009,Petit2005,Stauffer2017}. These cool gas clouds are seen in absorption, usually in the Balmer lines, when they transit the stellar disc. This is due to the scattering of chromospheric photons within the cloud. They can also be seen as an emission feature when they are beyond the stellar limb. So far, all the detections favour magnetically active cool stars (spectral types F,G and K) with rotation periods from $0.3-3$ days, although fast rotating M-dwarfs can also host these features. 

Stars that host prominences have ages ranging from $1$ Myr, such as the T Tauri star V410 Tau \citep{aaaaSkelly2010}, to $\sim 1$ Gyr as in the binary star QS Vir \citep{parsons2016} implying that this phenomenon may be present at different evolutionary stages. For an older solar-like star, we can assume that these prominences no longer form as they are not detected in our Sun, but they may have been present in the past. AB Dor is a solar analogue pre-main sequence star ($\sim 50$ Myr; \cite{Azulay2017}) showing a complex of prominences. \cite{collier1989I,collier1989II} found at least 6 prominences around AB Dor, distributed between $3$ and $9$ stellar radii. This places them well beyond the co-rotation radius ($2.6$ $r_\star$).  It is worth asking then over what phases of its evolution a star like our Sun would either start or stop forming prominences. 

In \cite{Villarreal2018} we classified the magnetosphere of a sample of cool-stars by their ability to form prominences. This classification depends on the value of two radii: the co-rotation radius, $r_\mathrm{K}=\sqrt[3]{GM_\star/\Omega^2}$, and the Alfvén radius, $r_\mathrm{A}$. Stars with $r_\mathrm{K}<r_\mathrm{A}$ may have a centrifugally-supported magnetosphere such that material flowing along closed magnetic field lines, beyond $r_\mathrm{K}$, can find a potential well where it will start to accumulate \citep{Jardine2001, Ferreira2000}. Mass accumulation in these potential minima will continue until the magnetic tension can no longer counteract the centrifugal force and the prominence is ejected.

The two radii mentioned above ($r_\mathrm{K}$ and $r_\mathrm{A}$) are dependent on the rotation rate of the star and it is expected that they will change as the star evolves. Stellar cluster observations of rotation period distributions at different ages have helped to reproduce the history of the rotational velocity of low-mass stars. Constrained by the observations, rotational evolution models \citep{Bouvier2014,gallet2015,Matt2015,Johnstone2015,Tu2015,Amard2016,See2018} have shown that during the pre-main sequence phase, low-mass stars can have a low rotation velocity, even when they are contracting and accreting material from the circumstellar disk. Once the disk is dissipated, the star spins up as it contracts onto the main sequence. Once on the main sequence, magnetized stellar winds carry angular momentum away from the star causing it to spin down. From the main sequence onwards, the star will continuously spin down. 

In this work we investigate the time span for which slingshot prominences may exist during the lifetime of a solar-like star i.e., for how long a star like our Sun satisfies the condition $r_\mathrm{K}<r_\mathrm{A}$, and also how the mass, mass-loss rate and lifetime of ejection of the prominence evolves. For this, we make use of the analytical expressions for the prominence parameters derived in our previous work \citep{Villarreal2018}. These parameters are found to be dependent on the co-rotation radius of the star, the stellar magnetic field, mass and radius and a geometric factor that depends on the assumed prominence's volume (taken from AB Dor's observations \citep{collier1990}). All of them are explicitly or implicitly  dependent of the rotational velocity of the star. The stellar rotation rate dependency of these parameters can be translated to a time dependency by means of a rotational evolution code. We adopt the rotational evolution code developed in \cite{Johnstone2015} and \cite{Tu2015} for a solar-like star from the pre-main sequence to the main sequence. 

In section \ref{method} we present the equations and the rotational evolution code that we used to obtain the time evolution for the prominence parameters. 
In section \ref{results} we explore the condition for the formation of prominences for a solar-like star and present our results and discussion. Conclusions are presented in section \ref{conclusion}.

\section{Methods}\label{method}
\subsection{Rotational Evolution Code}\label{REcode}

We employed the results from a rotational evolution code from the work of \cite{Johnstone2015} and \cite{Tu2015}. The code, originally developed to study the evolution of the stellar rotation and wind properties for low-mass stars on the main sequence \citep{Johnstone2015} is extended to the pre-main sequence in the work of \cite{Tu2015}. The free parameters in the model are constrained using measured rotation periods from several stellar cluster at different ages: NGC 6530 ($\sim 2$ Myr), h Per ($\sim 12$ Myr), Pleiades ($\sim 125$ Myr), M50 ($\sim 130$ Myr), M35 ($\sim 150$ Myr), NGC 2516 ($\sim 150$ Myr), M37 ($\sim 550$ Myr), Praesepe ($\sim 580$ Myr) and NGC 6811 ($\sim 1$ Gyr). Combining these rotation periods to get rotation period distributions at different ages (from $\sim2$ Myr to $\sim1000$ Myr) and binning the masses of the stars within these distributions, it is possible to get percentiles of the rotational distribution for every mass bin and age. In particular, the authors employed the $10$th, $50$th and $90$th percentiles of the distribution to fit their models.  

The formula that gives the variation of the stellar rotation with time involves a knowledge of the star's moment of inertia ($I_\star$) and the torque ($\tau_w$) on the star by the wind. 
To account for the variation of the moment of inertia, the radius,  and  the  convective  turnover  times  during  the  pre-main sequence, the model uses the stellar evolution model of \cite{spada2013}. To better match the observational constraints on rotation, the model uses the common assumption
(e.g. \cite{gallet2015}) that the interior of the star is not rotating with a single rotation rate, but instead has two separate rotation rates, one for the core and radiative zone and the other for the outer convective zone. The two regions exchange angular momentum with coupling timescales of $30$ Myr, $20$ Myr and $10$ Myr for the $10$th, $50$th and $90$th percentiles respectively. The model also assumes disk-locking timescales of $10$ Myr, $5$ Myr and $2$ Myr for the $10$th, $50$th and $90$th percentiles. During the main sequence the variation of $I_\star$ is negligible.

The wind torque is calculated using the formula derived by \cite{matt2012} and assuming that the dipole field strength of the star, $B_{\mathrm{dip}}$, and the wind mass loss rate, $\dot{M_\star}$, saturate at a Rossby number $Ro=0.13$, where $Ro=P_{\mathrm{rot}}/\tau_\star$ and $\tau_\star$ is the convective turn-over time. The scaling laws that determine the dependency of these parameters with $\Omega$ presented in the work of \cite{Johnstone2015} are written in terms of $Ro$ to take into account the change in $\tau_\star$ on the pre-main sequence \citep{Tu2015} and so  
\begin{align}
\dot{M}_\star &= \, \dot{M}_\odot \left(\frac{R_\star}{R_\odot}\right)^2 \left(\frac{Ro_\odot \tau_\odot }{Ro_\star \tau_\star } \right)^{a} \left(\frac{M_\star}{M_\odot} \right)^b, \label{eq:mdot}\\
B_{\star} &= \, B_{\odot} \left(\frac{Ro_\star}{Ro_\odot}\right)^{-1.32} .
\end{align}
The free parameters in Eq. \ref{eq:mdot}, $a$ and $b$, are calculated by fitting the model results with the observations. In our case $a=-2$ and $M_\star \sim M_\odot$. 
The assumed scaling law for the dipolar magnetic field, $B_\star$, is taken from the work of \cite{Vidotto2014} using the relation for the large scale averaged magnetic field, $<|B_{\rm{v}}|>$ with $\Omega$\footnote{We noticed an error in the exponent used in the paper of \cite{Tu2015}. When writing the scaling laws in terms of Ro, an exponent of 1.38 for the scaling of $B_\star$ should have been used in the model. This would not make any noticeable difference in our results and both exponent values falls within their errors.} since, as mentioned by Dr Vidotto in private communication, $<|B_{\rm{v}}|>$ scales linearly with $B_\star$.
For a more detail explanation of the origin of these scaling laws and the assumptions made in constructing them we refer the reader to the work of \cite{Johnstone2015} and \cite{Tu2015}.

\begin{figure}
\centering
\includegraphics[width=.48\textwidth]{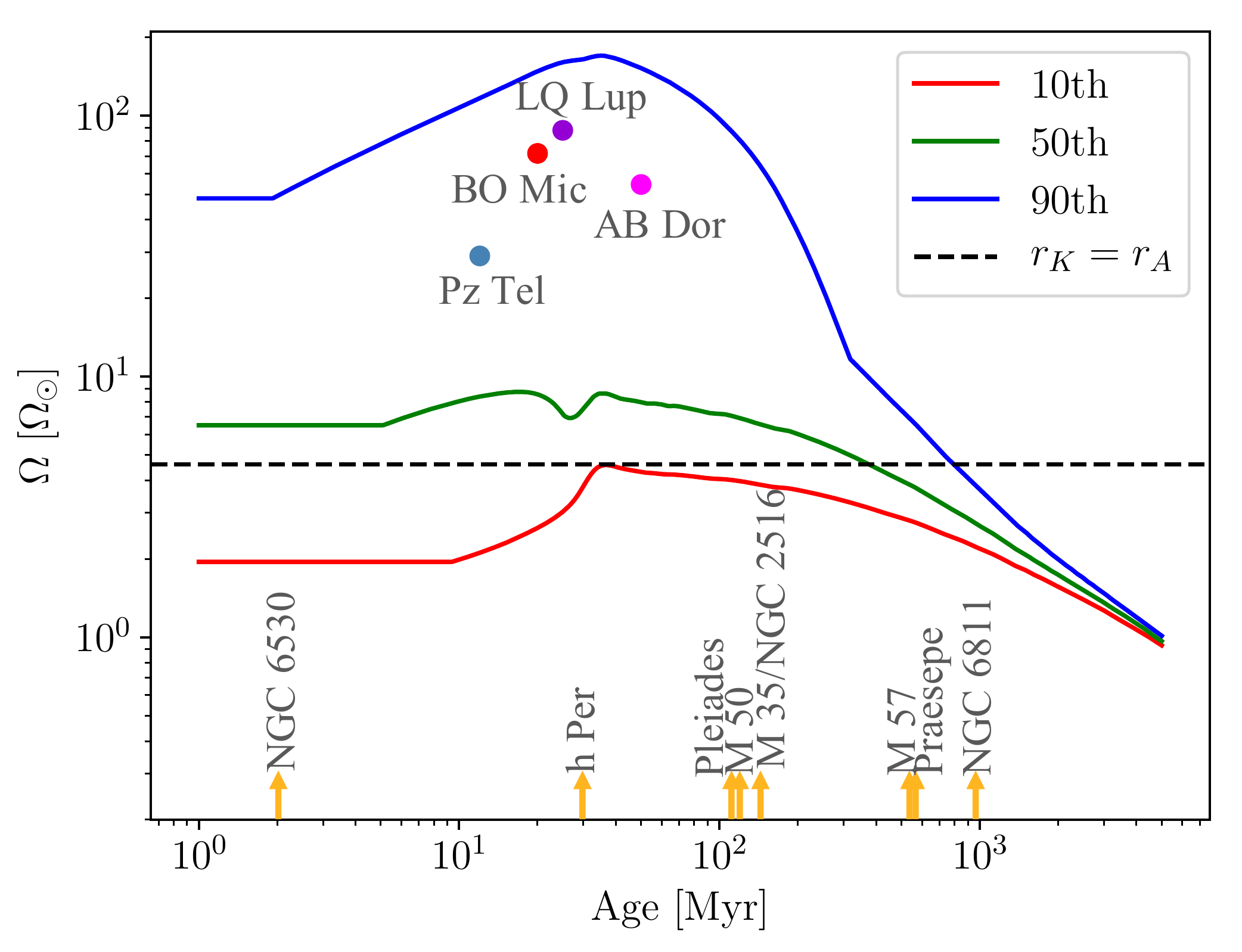}
\caption{Surface angular velocity as a function of age from the rotational evolution code \protect\citep{Tu2015}. Three different tracks corresponding to the $90$th percentile (blue) the $50$th percentile (green) and the $10$th percentile (red). The dashed black line indicates the angular velocity ($4.6\Omega_\odot$) for which $r_\mathrm{K}=r_\mathrm{A}$. Dots represent solar-like stars from Table \ref{tab:1} with masses around the $0.9$ - $1.1$ $M_\odot$ mass range.}
\label{fig:2}
\end{figure}

The predicted $\Omega$ evolution for stars in the $0.9$ - $1.1$ $M_\odot$ mass range for the $90$th, $50$th and $10$th percentiles of the observed rotational distribution are shown in Figure \ref{fig:2}. These tracks can be related to a fast, a medium and a slow rotator, respectively.
 The figure is the same as the one presented in the work of \cite{Tu2015}. The arrows on the x-axis show the ages of the clusters from where rotational period observations were measured.
 The initial velocities  in Fig. \ref{fig:2} are determined by the percentiles of the rotational distribution from the youngest cluster and they remain constant until the disk-locking period ends. 
The assumption of a constant stellar rotational velocity while the stellar disk is present has already been employed in previous rotational evolution models  \citep{Gallet2013,gallet2015}, and is based on observational results that favours an scenario of spin equilibrium during the first few Myrs in the life of the star \citep{Rebull2004,Herbst2005}. 
After  the disk is dissipated, the star begins to spin-up as it contracts towards the zero-age main sequence (ZAMS). Once the star reaches the ZAMS ($\sim 40$ Myr) it will start to spin down as a consequence of the loss of angular momentum mainly due to the magnetized stellar wind. Close to the age of the present Sun ($4.56$ Gyr) all the tracks converge to the same spin rate due to the strong dependence of the wind-braking mechanism with $\Omega$ \citep{Bouvier2014}.

\subsection{Prominence parameters}\label{equations}
The prominence parameters that we are interested in studying were already derived in our previous work \citep{Villarreal2018} and we briefly repeat here for completeness. 
Considering a dipolar magnetic field structure for the star, the prominence mass, mass loss rate and prominence lifetime are determined from the following set of equations: 
\begin{align}
m_\mathrm{p} &= \, \frac{B_\star^2r_\star^4}{GM_\star}\left(\frac{r_\star}{r_\mathrm{K}}\right)^2 F, \label{eq:1}\\
\dot{m}_\mathrm{p} &=  \,\frac{\dot{M}_{\star}}{4\pi r_{\star}^2} \, 2d A_{\star},\label{eq:2}\\
t_\mathrm{p} &=  \,\frac{m_\mathrm{p}}{\dot{m}_\mathrm{p}},\label{eq:3} 
\end{align}
where $G$ is the gravitational constant, $B_\star$ is the stellar average magnetic field value and $M_\star$ is the stellar mass. $r_\star$ is the stellar radius and $r_K$ the co-rotation radius. In Eq. \ref{eq:2}, $dA_\star$ is the area at the stellar surface that maps to the corresponding prominence area at the equator. Finally, the $F$ factor in Eq. \ref{eq:1} given by
\begin{equation}
F=\frac{\Delta \phi}{4\pi}\left[ \frac{1}{\bar{r}^2} - 
        \frac{1}{3}\ln\left(  \frac{1-\bar{r}^3}{(1-\bar{r})^2}\right) +
        \frac{2}{\sqrt{3}} \tan^{-1}\left(\frac{2\bar{r}+1}{\sqrt{3}} \right)
        \right]^{\bar{r}_{\rm max}}_{\bar{r}_{\rm min}}, \nonumber
\end{equation}
is a geometric factor that accounts for the integration in r and $\phi$, in terms of $\bar{r}=r/r_\mathrm{K}$. We remind the reader that this set of equations corresponds to a single prominence at the stellar equator with a volume determined by the extension in $\phi$ and $\bar{r}$, and with the assumption that the scale height of the prominence material is equal to the radius of curvature of the magnetic flux tube that contains it. 

All the stellar variables in Eq. \ref{eq:1}-\ref{eq:3} ($B_\star$, $M_\star$, $r_\star$ and $r_\mathrm{K}$) depend, directly or indirectly, on the stellar rotation and they will evolve as the star spin down through time. The change in these parameters with the rotation rate of the star can be translated to a time evolution employing the rotational evolution code presented in the previous subsection.

\section{Results \& Discussion}\label{results}
\subsection{At what angular velocity did our Sun stop forming slingshot prominences?}\label{when}
In our previous work \citep{Villarreal2018} we found, using the classification proposed by the massive star community \citep{Petit2013}, that for a star to been able to support slingshot prominences, the Alfvén radius should be larger than the co-rotation radius. Since both radii will change with the stellar rotation rate, the condition for supporting 
prominences ($r_\mathrm{K} < r_\mathrm{A}$) may not hold during the entire life of the star. Indeed, these types of prominences are not seen in our Sun at the present time.   

For a solar-like star, we can calculate how the co-rotation and Alfvén radii change with $\Omega$. Ideally, we would be interested in calculating the evolution of both radii during the entire lifetime of the star, but determining the Alfvén radius is not a easy task. It requires the knowledge of several stellar parameters currently not well constrained in the early stages of stellar evolution. We have therefore chosen to calculate both radii during the main sequence phase only. Although this approach does not allow us to constrain the age at which prominences may start to form in the pre-main sequence, it does give constraints on when a solar like star will stop forming prominences at later ages.

\begin{figure}
\centering
\includegraphics[width=.48\textwidth]{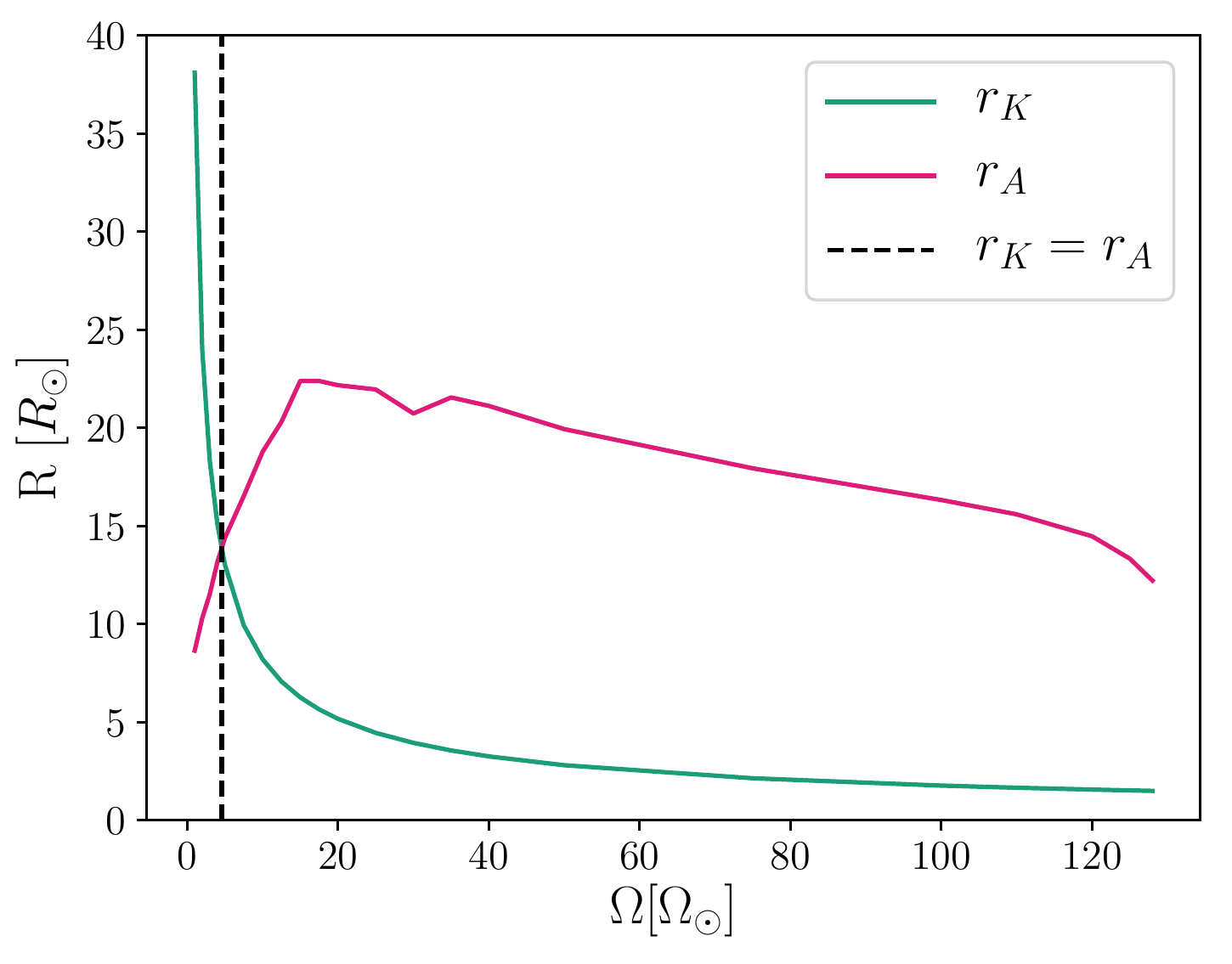}
\caption{Co-rotation radius and Alfvén radius as a function of $\Omega$ for a solar-like star. The dashed black line indicates the value of $\Omega_\mathrm{eq}=4.6\,\Omega_\odot$ where $r_\mathrm{K}=r_\mathrm{A}$.}
\label{fig:1}
\end{figure}

The $r_\mathrm{A}$ values are taken from Model A for a main sequence star of $1 \, M_\odot$ calculated in the work of \cite{Johnstone2017}. In this work, the authors employed a 1D MHD wind model calculated using the Versatile Advection Code (VAC; \cite{toth1996}) and computed the different stellar wind solutions that result when varying the input parameters as a function of $\Omega$ \citep{Johnstone2015b}. For every value of $\Omega$ we then have the corresponding Alfvén radius. The results are plotted in Fig. \ref{fig:1} together with the co-rotation radius as a function of the rotation rate for the same star.  

In Fig. \ref{fig:1}, the Alfvén radius and co-rotation radius are equal at $\Omega_{\mathrm{eq}}=4.6\,\Omega_\odot$. Therefore, a solar-like star can support prominences during the main-sequence phase until its angular velocity reaches $\Omega_\mathrm{eq}$. We also explored an alternative calculation of $r_\mathrm{A}$ using the formula presented in \cite{matt2012}. We found that using this prescription does not substantially change the estimated value of $\Omega_{\mathrm{eq}}$. In the pre-main sequence, due to the lack of a consistent calculation of the Alfvén radius, we assume that the star could begin forming prominences after the protoplanetary disc is dissipated, if its initial rotation rate is larger than $4.6\,\Omega_\odot$. The total period over which we expect the presence of prominences for this type of stars is showed as the shadow region in Fig. \ref{fig:3}.

Figure \ref{fig:2} also shows the value of $\Omega_{\mathrm{eq}}$ (dashed black line) for which the co-rotation and Alfvén radii become equal. The position where this line crosses the three evolutionary tracks give us the age at which a solar-like star would stop forming prominences as the condition $r_\mathrm{K}\leq r_\mathrm{A}$ no longer holds. For a fast and medium rotator ($90$th and $50$th percentiles) the ages are $\sim 800$ and $\sim 400$ Myr respectively. In the case of a slow rotator ($10$th percentile) the $\Omega_{\mathrm{eq}}$ line never crosses the evolutionary track, suggesting that a slow rotator may never form this type of prominences during its life.

\subsection{Evolution for a single slingshot prominence in a solar-like star}

\begin{figure}
\centering
\includegraphics[width=.5\textwidth]{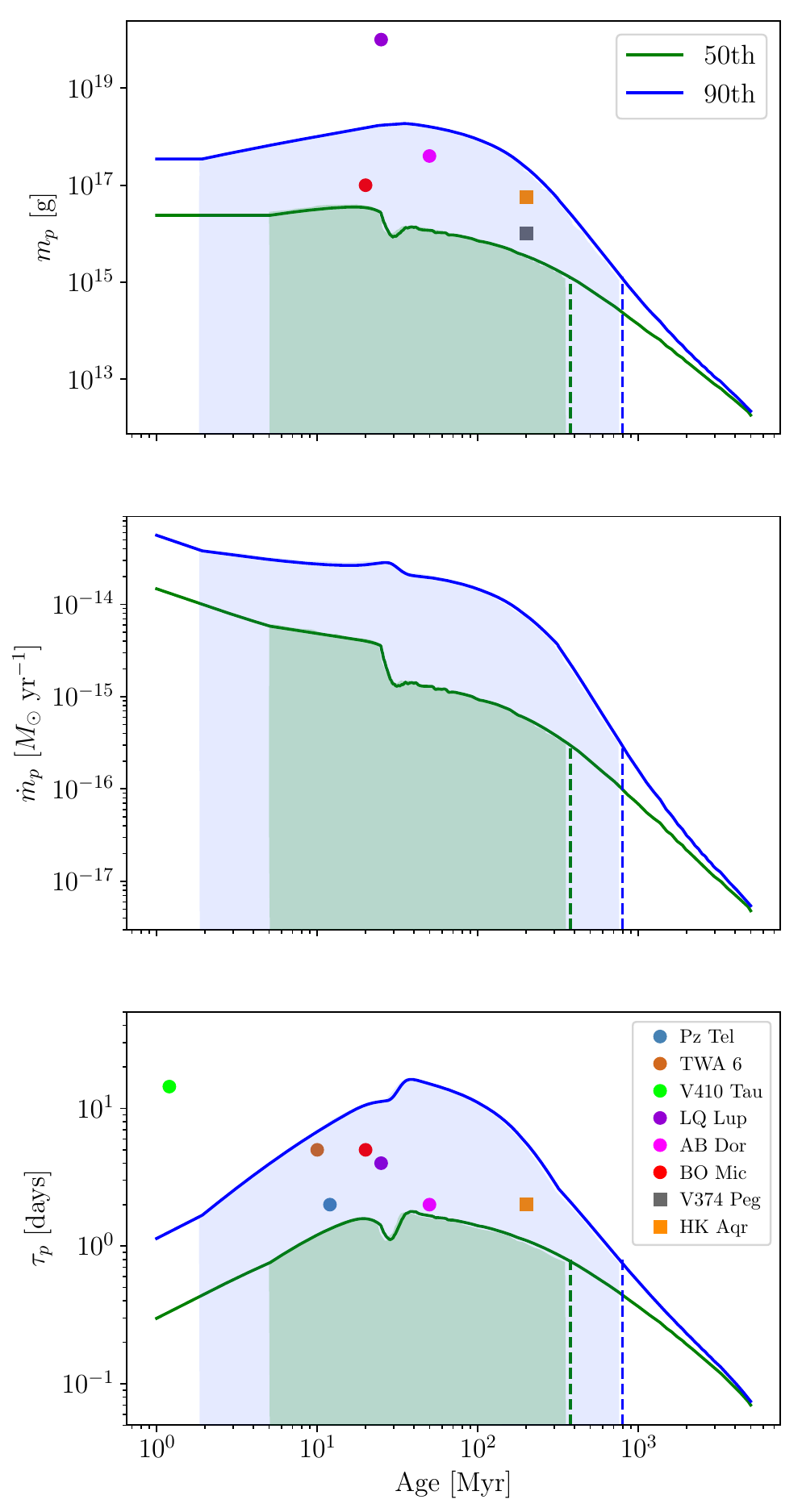}
\caption{Prominence mass (top), mass loss rate (middle) and lifetime (bottom) as a function of stellar age for a solar-like star, for two tracks corresponding to the fast (blue) and medium (green) rotator ($90$th and $50$th percentile). The dashed vertical lines correspond to the ages at which the star can no longer support prominence formation, $\sim 800$ Myr for a fast rotator and $ \sim 400$ Myr for a medium rotator. The shaded area represents the estimated time in the stars life for which prominences are present.}
\label{fig:3}
\end{figure}

%---------Table 1-----------------------------------------------------
\begin{table}
    \setlength\tabcolsep{2.5pt}
	\caption{Stellar parameters and observed prominences characteristics.}
	\label{tab:1}
    \begin{threeparttable}	
        \noindent \begin{tabular}{lccccccc}
	    \hline
        \hline
        Name & Mass       &age  & S. Type & $P_\mathrm{rot}$& $m_\mathrm{p}$ & $t_\mathrm{p}$& Ref.\\
             &[$M_\odot$] &[Myr]&         & [d]      & [g]   & [d]  &     \\
        \hline
        \textbf{Cool stars} &&&&&&&\\
        PZ Tel    & 1.13 & 12 & G6.5  & 0.94 & -          & <2    & (1) \\
        BO Mic    & 0.82 & 20 & K3V   & 0.38 & $10^{17}$ & >5    & (2) \\
        AB Dor	  & 0.86 & 50 & K2V   & 0.5  & $4\times10^{17}$   & >2    & (3) \\
        TWA 6     & 0.7  & 10 & K7    & 0.54 & -          & <5    & (4) \\
        V410 Tau  & 1.4  & 1.2  & K3    & 1.871& -        & >14.4 & (5) \\
        LQ Lup    & 1.16 & 25 & G8IVe & 0.31 & $10^{20}$  & <4    & (6) \\
        %V471 Tau  & 0.93 & 600& K2V   & 0.52 & (0.1-1)e11 & -    & (7) \\
        \hline
        \textbf{M-dwarfs} &&&&&&&\\
       % QS Vir    & 0.382 & 1000& dMe  & 0.1 & - & - & \\
        HK Aqr   & 0.4  & 200& dM1.5e & 0.43 & $5.7\times10^{16}$ & > 2 & (7) \\
        V374 Peg & 0.3  & 200& dM4    & 0.44 & >$10^{16}$         & -   & (8) \\
        \hline
        \hline 
   \end{tabular}  
   \begin{tablenotes}[normal,flushleft,online]
    (1) \protect\cite{Barnes2000,Leitzinger2016}. (2) \protect\cite{Jeffries1993,Dunstone2006,Dunstone2006b}. (3) \protect\cite{collier1989I,collier1989II,collier1990}.  (4) \protect\cite{Skelly2008}. (5) \protect\cite{aaaaSkelly2010}. (6) \protect\cite{Donati2000}. 
    %(7) \protect\cite{Guinan1986}.
    (7) \protect\cite{Byrne1996,Leitzinger2016}. (8) \protect\cite{vida2016}.
  \end{tablenotes}
\end{threeparttable}
\end{table}
%------------------------------------------------------------
The stellar parameters as a function of the age obtained with the rotation evolution model can be introduced in Equations \ref{eq:1}-\ref{eq:3} to get the mass ($m_\mathrm{p}$), mass loss rate ($\dot{m}_\mathrm{p}$) and lifetime ($\tau_\mathrm{p}$) of a single prominence as a function of age, for a star of $1 M_\odot$. The estimation of these parameters needs the assumption of a prominence volume, which we have taken to be the one derived for AB Dor \cite{collier1990}. 
The results are shown in Figure \ref{fig:3} for the $90$th (fast rotator) and $50$th (medium rotator) percentiles. The shadowed area in the figures represents the time during the life of the star when prominences can form. Even though we lack of a consistent calculation of $r_\mathrm{A}$ in the pre-main sequence, the prominence parameters can be calculated in this regime as the do not depend on $r_\mathrm{A}$.
From Fig. \ref{fig:3} we conclude that as the star evolves, the mass and ejection time follow a similar trend to the one observed for $\Omega$, with maximum values found closer to the time when the star reaches the ZAMS. Prominence masses at this stage reach $10^{18}$ and $10^{16}$ g for the fast and medium rotators with ejection times of $1$ to $14$ days respectively.
Mass loss rates for the prominences show a more constant behaviour during the pre-main sequence phase, decreasing towards the main sequence. 

In this paper we have used one particular rotational evolution model to illustrate the age span over which slingshot prominences may be expected to form. While the quantitative results may change with the use of another model, we expect the general trends to be robust.

Figure \ref{fig:3} also shows the observationally-derived values of mass and lifetime of prominences found in the list of stars presented in Table \ref{tab:1}. Circles represent cool stars while squares represents M-dwarfs. Most of the stars in Table \ref{tab:1} have a lifetime estimation for the prominences, but only a few of them have a mass estimation. The highest mass value estimated for slingshot prominences is found in LQ lup (G8IVe). In this case, and due to the fact that the star has a low inclination value ($i=35^\mathrm{o}$), the entire prominence complex can be observe in emission \citep{Donati2000} and so the mass value represents the total mass of the prominence system. In the other cases, the estimated masses represents a lower limit since they correspond to a single prominence, as in our model, or a few prominences present in the hemisphere that faces the observer.

In general, the observed values of $m_\mathrm{p}$ and $\tau_\mathrm{p}$ found for stars with masses close to our Sun in Fig. \ref{fig:3}, lie within the fast rotating track and they cover most of the time that our model predicts slingshot prominences should exist.
Additionally, we show the angular velocity and age of these stars on Fig. \ref{fig:2}. As expected, they fall between the medium and the fast rotating track of the rotational evolution model. This is not surprising as slingshot prominences were primary detected among fast rotators.

We are aware that values from the M-dwarfs and the very young star V410 Tau should not be included in this conclusion since they are not well represented by our model, because it considers only solar-mass stars. These stars will follow different evolutionary tracks from the one predicted by the model but they are shown on Fig. \ref{fig:3} for illustrative purposes. 

\section{conclusion}\label{conclusion}
We have employed a rotation evolution code to obtain stellar parameters as a function of age and coupled this with the prominence model developed in a previous work to calculate prominence masses, mass loss rates and lifetimes during the life of a solar-like star. 

We have found that the condition that enables the star to support the accumulation of material along closed magnetic field lines ($r_\mathrm{K} \leq r_\mathrm{A}$) can hold from the early stages of stellar evolution and will last a few million years after the star has entered the main sequence. For a fast rotator this age is around $800$ Myr, for a medium rotator it will be around  $400$ Myr, but for a star that begins with a low angular velocity (slow rotator), slingshot prominences may never form.

We have also found that for a medium and a fast rotator, prominences will be present for a considerable amount of time during their life. We believe that observing and characterising slingshot prominences at different stellar ages will help us to study the evolution of the coronal magnetic field. So far, studies of the structure of stellar coronae have been undertaken with the help of radio and X-ray emission observations for a few stars \citep[and references therein]{Hussain2007}. Detecting slingshot prominences is a much cheaper task since they can be observed as a transit phenomenon in the hydrogen lines.    

The detection of slingshot prominences in a single star is most favourable for prominences with higher masses and longer lives, which produce a more noticeable, longer-lived transit feature. According to our study, these requirements are met around $40$ Myr, when a solar-like star reaches the ZAMS.
It is interesting to point out, that the first example of slingshot prominence detection was obtained for AB Dor, which is a solar-like star with an age around $50$ Myr. Since that first detection, many young clusters have already been subject to spectroscopy studies, expanding the range of cluster ages within which such prominences may be detected.   

\section*{Acknowledgements}
We thank the referee for useful comments that helped improve this manuscript. CVD and MJ acknowledges STFC grant (ST/M001296/1). CVD acknowledges the Irish Research Council (IRC) postdoctoral fellowship (Project 208066, Award 15350). CPJ acknowledges the support of the FWF (Austrian Science Foundation) NFN project S11601-N16, and the related FWF NFN subproject S11604-N16. VS acknowledges funding from the European Research Council (ERC) under the European Unions Horizon 2020 research and innovation programme (grant agreement No 682393) AWESoMeStars.

\bibliographystyle{mnras}
\bibliography{sample}

\end{document}